# Performance Evaluation and Optimization of Air-to-Air Thermoelectric Heat Pump System

Je-Hyeong Bahk, Thiraj D. Mohankumar, Abhishek Saini, and Sarah J. Watzman

*Department of Mechanical and Materials Engineering, University of Cincinnati, Cincinnati, OH 45221, USA*

**Corresponding authors:**

J.-H. Bahk (bahkjg@ucmail.uc.edu), S. J. Watzman (watzmasj@ucmail.uc.edu)

## Abstract

Thermoelectric (TE) heat pumps are a promising solid-state technology for space cooling and heating owing to their unique advantages such as environmental friendliness with no harmful refrigerants, small form factors, low noise, robustness with no moving parts, and demand-flexible operation. However, their low coefficient of performance (COP) presents a challenge towards broader adaptation of the technology at the market level. Built on our previous theoretical framework for a modular system design aimed at improving both COP and cooling capacity, this study focuses on experimental characterization and optimization of a unit air-to-air TE heat pump system. We employ double-sided heating/cooling and counterflow configuration for uniform heat transfer with TE modules and heat exchangers. System-level evaluations validate the theoretical model and show the variation of air temperature change and system COP with different electric current inputs and air flowrates. A maximum cooling COP of 4.4 and a maximum heating COP of 6.0 were obtained at optimized current levels. Further analysis shows that parasitic thermal resistances in the system significantly reduced both COP and air temperature change by over 50%.

# 1. Introduction

Global energy consumption in residential and commercial buildings reached 133 EJ in 2022, accounting for approximately 25 % of the world's total energy demand, and is expected to increase to 160 EJ by 2050. [1] Among the various uses of energy in buildings, space heating and cooling represent about a half of the total energy consumption. Improving energy efficiency in space conditioning can have major impacts on the energy costs in homes and commercial buildings. Conventional vapor-compression heat pump systems rely on the use of refrigerants that are toxic and potentially causing climate change and suffer from excessive noise during the compression/expansion of refrigerant, limited energy efficiencies, and large space occupancies.

Thermoelectric (TE) energy conversion has attracted great attention in recent years because of its ability to directly convert heat to electricity or vice versa using a simple solid-state semiconductor device offering sustainable solution in various fields. [2] TE heat pumps have many advantages such as being environmentally friendly with no refrigerant use, compactness, simple construction, noiseless operation, robustness without moving parts, and demand flexible operation due to their electrically controlled working principle. Both cooling and heating can be achieved based on the Peltier effects, in which heat is pumped from one side of the device, thus cooling, and transferred to the other side, thus heating, by supplied electric current. Simply flipping the electric current direction can switch the sides of cooling and heating. A heat exchanger can be attached at the cooling side of the TE module, through which supply air flows and gets cooled. The other side of the TE module can be either water-cooled or air-cooled to remove the pumped heat efficiently. [3] Although water cooling can achieve a higher TE cooling efficiency, it might make the system more complex with additional water pipelines and consume extra water. Instead, air-to-air TE pumps can eliminate the use of water by dumping heat to exhaust air that already exists in the building.

Recently, there have been many studies on TE air cooling and heating technologies. For water-cooled systems, Fan et al. [4] proposed a method to analyze water-cooled TE air cooling systems, validated by experimental data. They predicted a maximum cooling density of 8.65 kW/m² and an optimal coefficient of performance (COP) of 2.27 at a 5 K cooling, while achieving experimental COP of 0.95. Ma et al. [5] reviewed TE air conditioning (A/C) technologies and emphasized the need for better analytical models and improved TE materials. Duan et al. [6] developed a validated mathematical model for TE air conditioning systems, highlighting the efficiency boost possible with optimized thermoelectric materials, suggesting a *ZT* value of 3 could surpass traditional heat pumps. Mao et al. [7] provided a review on TE cooling materials. Looi et al. [8] demonstrated a photovoltaic TE A/C system achieving an average COP of 1.67 and reducing indoor temperature by 4 °C. Liu et al. [9] explored a solar thermoelectric system, achieving a COP of 0.9 in cooling and 1.9 in heating. Dizaji et al. [10] enhanced a TE A/C device's COP to through module jointing and heat sink improvements in air-to-water cooling system, citing benefits of low voltage. Ibañez-Puy et al. [11] tested bi-directional thermoelectric devices, achieving a cooling COP of 2.8 and a heating COP of 2.2. Yilmazoglu [12] noted a TE cooler's performance impact by voltage and airflow, with a cooling

COP under 1 and a heating COP of 2.5 ~ 5. Liu et al. [13] attained a COP of 4.5 in cooling mode with water cooling, observing efficiency drops at higher water temperatures. Cosnier et al. [14] achieved a cooling COP of 1.5 and heating COP of 2 by maximizing electrical intensity between 4-5 A and minimizing temperature differences across TE modules.

There also have been several air-to-air TE systems reported in recent years. Attar et al. [15] designed an air-air TE cooling system, obtaining a COP of 2.1 for air temperature drop of 0.5 K and showing the trade-off relationship between COP and the temperature drop. Irshad et al. [16] explored a TE A/C system integrated in air ducts, achieving COP between 0.4 and 0.68 for temperature difference 3 ~ 5 °C, and cooling load of 499 W. Shen et al. [17] developed a system using TE modules as radiant panels, attaining a peak COP of 1.77. Hou et al. [18] showed that a TE heat pump can achieve a COP of up to 3.7 increased by 50~ 70 % with air recirculation and regenerative heat recovery. Erro et al. [19] investigated TE heating modules to preheat airflow for thermal energy storage and achieved COP ~ 4 to achieve ~ 4 K temperature rise at 25 °C inlet temperature.

In our previous theoretical investigation [3], we showed that there is a trade-off relationship between the COP and the temperature drop in a TE cooling system. A high-COP cooling can be achieved only with small cooling capacity or small temperature drop in air. To overcome this limitation, a sequential cooling scheme was proposed to achieve a large temperature drop by installing multiple TE modules in series along the airflow direction to continuously cool the air until the desired temperature is reached. The COPs of individual modules will remain high with distributed cooling power over multiple modules, and thus a high COP for the whole system. Also found is that the parasitic thermal resistances created at the interfaces between the TE module and the heat exchanger can significantly affect the cooling performance. Accurate quantification of these parasitic resistances is key to optimizing the performance and achieving a high COP.

In this work, we combine theoretical and experimental studies with a unit air-to-air TE heat pump system to evaluate the COP and temperature change of supply air through the system. We investigate both air cooling and heating performance with varying electric current and air flowrate. We also estimate the parasitic thermal resistance at the interface between TE module and the heat exchanger by curve-fitting the experimental data with the theory and finally discuss their impacts on the cooling and heating performance.

## 2. Experimental Methods

In our experimental setup, we employ a double-sided TE heat pumping scheme for air-to-air heat transfer with counter-flow configuration. Fig. 1 shows the schematic flow configuration with the TE module and heat exchanger assembly, and a picture of the complete setup built in this work. As shown in Fig. 1(a), supply air passes through the middle duct and gets cooled or heated by two TE modules from both sides. The double-sided TE heating and cooling ensures uniform heat transfer from or to the supply air even at high flowrates. For efficient heat transfer, heat exchangers are integrated in all the three air ducts. Simple plate-fin heat exchangers made of aluminum alloy (AA 6063) are used in our system for minimal pressure drop and high heat transfer rate.

In the cooling mode schematically shown in Fig. 1(a), the TE modules pump heat out of the supply air flowing in the middle duct and transfer the heat vertically to the heat exchangers in the top and bottom ducts. Exhaust air passing through these outer ducts absorbs the transferred heat from TE modules and carries heat to outside. In the heating mode shown in Fig. 1(b), the direction of the electric current supplied to the TE modules is flipped and Peltier heating occurs at the supply air side of the module. The supply air is heated up, and the exhaust air in the outer ducts is cooled down instead. Our TE system is designed to be modular. The prototype system shown in Fig. 1 shows only one repeat unit. Multiple parallel units can be added side by side to scale up the total flowrate. Multiple units can also be placed one after another in series in the flow direction for sequential cooling to scale up the cooling/heating power while maintaining the system COP high.

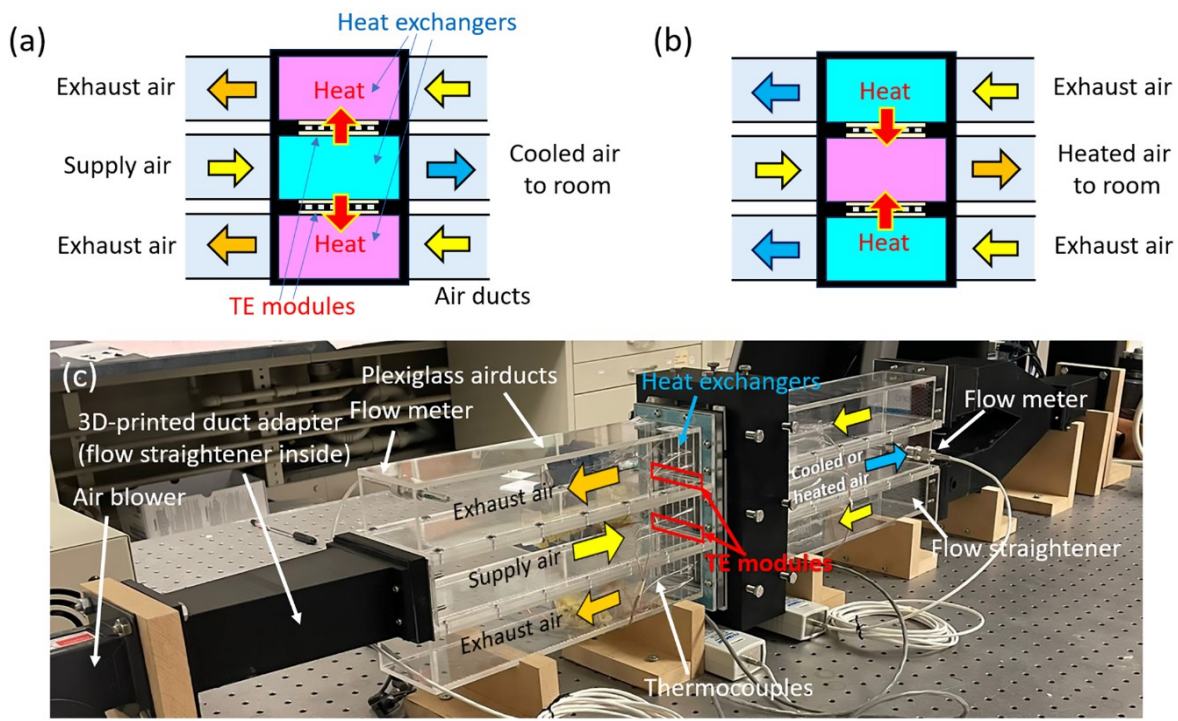

**Fig. 1.** (a)-(b) Schematic diagrams of the air-to-air TE heat pump system in counter-flow configuration with three air ducts – the middle duct for supply air flowing from left to right, and the two outer ducts for exhaust air flowing from right to left. (a) shows heat transfer in cooling mode and (b) in heating mode. (c) A picture of the complete system. The black assembly in the middle houses one repeat unit made of two TE modules sandwiched by three heat exchangers.

The heat exchanger-TE module assembly shown in the middle of the system was specially designed to accommodate different heat exchanger types. The core section of heat exchangers is 70 × 80 × 58 (W × L × H) mm$^3$ in size. In this work, we used plate-fin heat exchangers with fin thickness 4.5 mm, fin spacing 6 mm, and fin height 46.7 mm, made with aluminum alloy AL-6063, of which thermal conductivity is ~210 W/mK. Commercial TE modules (Marlow, PL080-8.5-40) were selected to use in our system because their design (fill factor and leg thickness) was close to the optimal design we found from our previous theoretical study [3]. Note that the TE module was smaller (40 × 40 mm$^2$) than the heat exchanger area (70 × 80 mm$^2$). There can be significant heat spreading or contraction when heat flows from TE module to heat exchanger, which adds additional thermal resistance. This heat spreading/contraction effect is included in our parasitic thermal resistance estimate as discussed later in this paper. A high-conductivity graphite pad (Panasonic EYG-TF0F0A05A, thickness 0.5 mm) was used as the thermal interface material between the TE module and heat exchanger. They were assembled with fastening screws to exert enough pressure to the graphite pad at the interfaces for minimum interface resistance and maximum heat transfer. All the sides of the assembly were insulated with an aerogel composite (Cryogel Z blanket) before encasing it with the 3D-printed black cover.

Temperatures at the base of heat exchangers and at the hot and cold sides of TE modules were measured using K-type thermocouples. Air temperature difference between inlet and outlet of each duct was directly measured using the differential temperature measurement technique. Two identical thermocouples were connected with each other to the opposite polarities and the two ends of the connected thermocouples were placed near the inlet and outlet of the heat exchanger to measure the temperature difference of air before and after the heat exchanger. The temperature difference was measured at four different places near each inlet and outlet and then averaged over the four measurements to minimize random errors and variability inherent in single measurements for more reliable data. Room temperature was measured to be 23.2 °C. Relative humidity in the room was found to be less than 20 %, which indicates that the moisture in air has negligible effects in the sensible cooling.

Air mass flowrate was set between 1.5 g/s and 10 g/s using a variable-flow blower. It is noted that this flowrate range is relatively low compared to those of standard air conditioning systems. Small-scale air conditioning systems typically require about 1 c.f.m. flowrate per square footage, which corresponds to ~0.58 g/s per square foot. In this study, flow rates up to 10g/s were examined, which corresponds to cooling a space up to around 17 ft$^2$. This flowrate, however, can be easily scaled up by having multiple parallel units for a larger space. The exact flowrates at both supply air and exhaust air sides were measured using hot-wire anemometer-type flowmeters (Omega FMA900A). The flowmeters were positioned five duct diameters downstream from the heat exchangers to ensure accurate measurements. All temperature and flow rate measurements were collected by a data acquisition system (National Instruments NI-9214, NI-9215 and cDAQ9174) and recorded in real time by a LabVIEW program. For each set of measurements, it was ensured that the convection conditions were stable, and steady-state temperatures were reached across all the measurements to obtain the final experimental data.

By conservation of energy, the total rate of heat pumping by the TE modules at the supply air side must be equal to the total internal energy change rate in the supply air, which is obtained by measuring the change in bulk temperature of the supply air such that

$$Q_{pumped} = \dot{m}_{sa} c_p \Delta T_{sa} \tag{1}$$

where $\dot{m}_{sa}$ is the mass flowrate of the supply air, $c_p$ is the specific heat of the supply air, and $\Delta T_{sa} = T_{sa,out} - T_{sa,in}$ is the change in bulk temperature of the supply air from inlet to outlet. We used the air properties at room temperature (~23 °C) under atmospheric pressure in this calculation. [3] The two TE modules are electrically connected in series to supply the same electric current with one voltage supply. The total electric power input is obtained as

$$P_{in} = V_{sup} I \tag{2}$$

where $V_{sup}$ is the total supply voltage, and $I$ is the electric current input to the TE modules. The coefficient of performance is then obtained by taking the ratio between the total pumped heat and the electric power input as

$$COP = \frac{Q_{pumped}}{P_{in}} \tag{3}$$

Material properties of the TE modules were obtained by impedance spectroscopy measurements, following the original work by García-Cañadas and Min [20] for TE module characterization under suspended conditions. Fig. 2 shows the resulting Nyquist plot for a TE module. The extracted material properties are summarized in Table 1. Note that the material properties are the average values between the multiple p-type and n-type TE elements in the TE module.

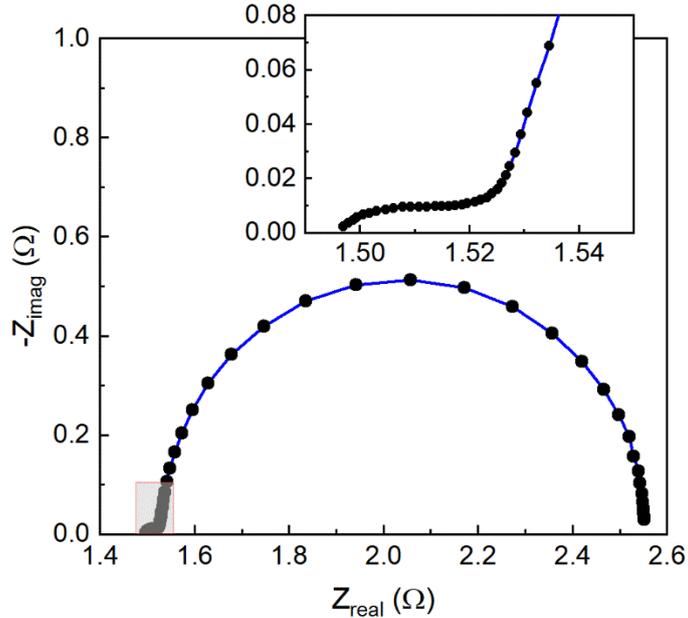

**Fig. 2.** Nyquist plot for the TE module being suspended in air. Inset shows a zoom-in view of the high frequency responses within the shaded area in the plot. This plot was used to extract material properties according to *[20]*.

**Table 1.** Material properties obtained from impedance spectroscopy analysis and the geometric properties for the TE module

| Seebeck coefficient, $S$ (μV/K) | Thermal conductivity, $k$ (W/m·K) | Electrical conductivity, $\sigma$ $(\Omega\cdot m)^{-1}$ | $ZT$ | Module area (mm²) | Leg area, $A_{TE}$ (mm²) | Leg length, $L_{TE}$ (mm) | No of legs, $N$ | Fill factor |
|---|---|---|---|---|---|---|---|---|
| 196.4 | 1.37 | $1.19 \times 10^5$ | 0.97 | 40 × 40 | 2.25 | 1.1 | 256 | 0.36 |

## 3. Theoretical modeling

Since our TE system is completely symmetric around the center line of the supply air duct, we modeled only one half of the system below the center line as schematically shown in Fig. 3(a). Only a half height of the supply air-side heat exchanger is included in the model, while the exhaust air-side heat exchanger is considered at its full height. Both far top and bottom sides of the model are assumed to be perfectly insulated. In the cooling mode, supply air flows through the heat exchanger from left to right, during which it is cooled by the TE module, losing heat, $Q_{sa}$, to the TE module. Here,

$$Q_{sa} = \frac{1}{2} Q_{pumped} \qquad (4)$$

because the model describes a half system only. As a result of the cooling, the supply air temperature decreases from $T_{sa,in}$ to $T_{sa,out}$. At the other side of TE module, $Q_{ea}$ is dissipated to the exhaust air through the bottom heat exchanger. This heat increases the exhaust air temperature from $T_{ea,in}$ to $T_{ea,out}$ as it flows from right to left.

We model the vertical heat transfer through the heat exchanger-TE module assembly as a 1D heat flow through multiple thermal resistances in series as described in Fig. 3(b). The Peltier and Joule heat components that are created in the TE module are modeled as heat current sources at the two ends of TE legs. Note that the direction of the Peltier heat at the supply air side, $Q_{P,sa}$, shown in Fig. 3(b) is away from the node, indicating heat absorption or cooling of the node, whereas the direction of the Peltier heat at the exhaust air side, $Q_{P,ea}$ is into the node, indicating heating of the node. In the heating mode, these directions are reversed. The magnitudes of $Q_{P,sa}$ and $Q_{P,ea}$ are different from each other, because they depend on the temperature at their locations ($T_{TE,sa}$ and $T_{TE,ea}$), i.e., $Q_{P,sa} = NST_{TE,sa}I$ and $Q_{P,ea} = NST_{TE,ea}I$, where $N$ is the number of TE legs, and $S$ is the average Seebeck coefficient of TE legs.

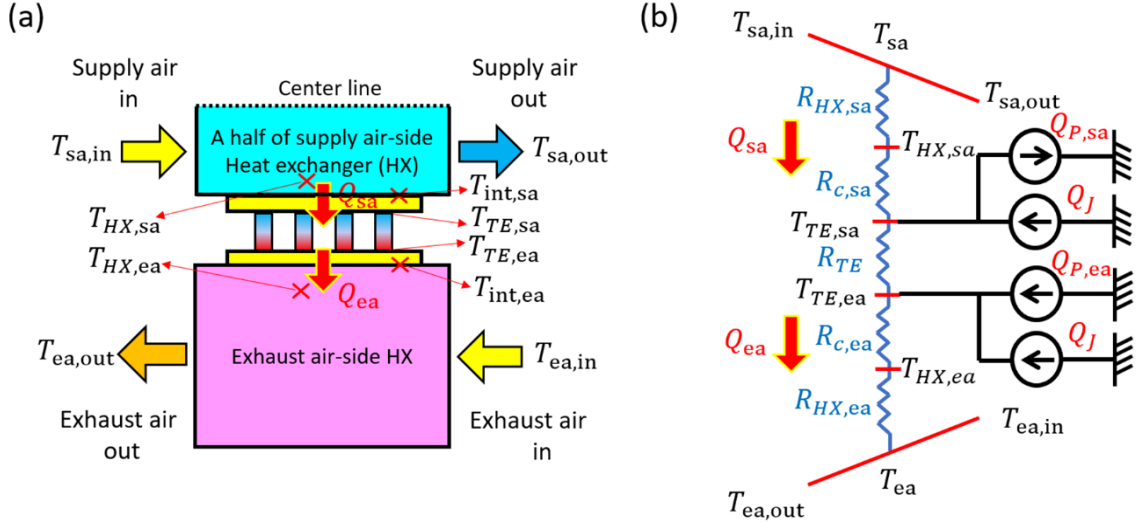

**Fig. 3.** (a) Schematic of the half-system model. The × marks indicate the locations where temperatures were measured experimentally. (b) One-dimensional thermal circuit model representing the half system. Peltier and Joule heat components created in the TE module are shown as current sources with subscripts, $P$ and $J$, respectively. Subscripts, sa and ea, indicate properties at the supply air and ehaust air sides, respectively. The directions of all heats shown in the drawing are in the case of cooling mode. In the heating mode, the directions of $Q_{sa}$, $Q_{ea}$, $Q_{P,sa}$ and $Q_{P,ea}$ are reversed, while the direction of $Q_J$ remains the same.

Joule heating that occurs inside the TE legs are equally divided and dissipated to the two ends, such that $Q_J = \frac{1}{2}I^2 R_e$, where $R_e = \frac{NL_{TE}}{\sigma A_{TE}} + 2NR_{e,c}$ is the total electrical resistance of the TE module that includes the TE leg resistances (the first term) and the contact electrical resistances (the second term). The latter is found to be 45 mΩ after measuring the total electrical resistance of the module and subtracting the calculated TE leg resistances with the material properties and dimensions from it. At the far two ends of the 1D resistance network are the two air temperatures, which are the average temperature of the inlet and outlet temperature of each air, i.e., $T_{sa} = (T_{sa,in} + T_{sa,out})/2$, and $T_{ea} = (T_{ea,in} + T_{ea,out})/2$.

As shown in Fig. 3(b), there are five thermal resistances in series in this resistance network model. $R_{HX,sa}$ and $R_{HX,ea}$ are the total heat exchanger resistances at the supply air side and the exhaust air side, respectively, between the air temperature ($T_{sa}$ or $T_{ea}$) and the heat exchanger base temperature ($T_{HX,sa}$ or $T_{HX,ea}$). They include all the fin resistances combining conduction through the fins and convection over the fin surfaces, as well as the convection resistance from the portion of the base plate area that is exposed to the air flow in between fins. [21] $R_{TE}$ is the total thermal resistance by the parallel TE legs in the TE module, given by $R_{TE,leg} = \frac{L_{TE}}{NkA_{TE}}$ as well as the parallel resistance through the filler (air) in between TE legs. The detailed discussion on the parasitic filler resistance is found in our previous paper. [3]

$R_{c,sa}$ and $R_{c,ea}$ are the total interface thermal resistances at the supply air side and the exhaust air side, respectively, between the heat exchanger base temperature ($T_{HX,sa}$ or $T_{HX,ea}$) and the TE end-leg temperature ($T_{TE,sa}$ or $T_{TE,ea}$). They are the sum of the conduction resistances from the heat exchanger base plate, the interface between the heat exchanger and the TE module made of the graphite pad, and the ceramic plate of the TE module. In particular, the resistances of the heat exchanger base plate and the TE ceramic plate include heat spreading/shrinking effects due to the difference in area between the two plates. It is important to note that $R_{c,sa}$ and $R_{c,ea}$ are not experimentally determined because we were not able to directly measure their boundary temperatures, $T_{TE,sa}$ or $T_{TE,ea}$, inside the TE modules due to the small thickness of TE legs. Instead, we determined $R_{c,sa}$ and $R_{c,ea}$ by curve-fitting the experimental temperatures and COPs with our theoretical model.

At the supply-air side end of TE legs inside the TE module, i.e., at the node marked by temperature $T_{TE,sa}$ in Fig. 3(b), the pumped heat from supply air, $Q_{sa}$, must be balanced with the Peltier and Joule heats at that node, $Q_{P,sa}$ and $Q_J$, as well as the conduction heat through $R_{TE}$, $Q_K = \frac{(T_{TE,ea} - T_{TE,sa})}{R_{TE}}$, due to the temperature difference between $T_{TE,sa}$ and $T_{TE,ea}$, considering their directions:

$$Q_{sa} = Q_{P,sa} - Q_J - Q_K \tag{5}$$

Similarly, at the exhaust-air side of TE legs, the heat balance equation reduces to

$$Q_{ea} = Q_{P,ea} + Q_J - Q_K \tag{6}$$

More detailed information about our modeling is found in our previous paper. [3]

## 4. Results and Discussion

In our experiments, we varied the mass flowrate of supply air from 1.5 to 10 g/s and obtained the temperature change in supply air from inlet to outlet and the resulting COP as a function of current from 0.5 to 3 A at each flowrate. Both cooling and heating were tested. We used the two interface thermal resistances, $R_{c,sa}$ and $R_{c,ea}$, defined in the theory section as fitting parameters to fit all these experimental data. The obtained values are $R_{c,sa}$ = 0.16 K/W, and $R_{c,ea}$ = 0.15 K/W. The two values are slightly different from each other potentially due to the small difference in the contact resistance by the interface graphite pads at both sides, but they are reasonably close to each other, as expected.

Fig. 4 shows example transient responses of supply air by the TE heat pump in cooling mode. As discussed in the experiment section, we directly measured the temperature difference between the inlet and outlet of supply air, instead of measuring the two temperatures separately. Due to the small flowrates used, the temperature difference between the inlet and outlet of supply air reaches steady state quickly in 2 ~ 3 seconds for the two flowrate 3 and 5 g/s, respectively. The time constants are found to be similar, 1.1 ~ 1.2 seconds, for both flowrates.

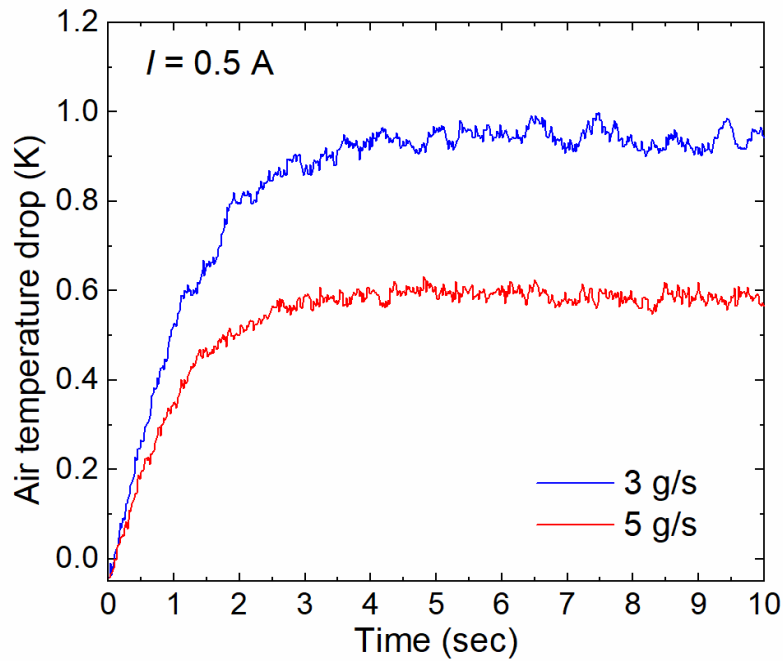

**Fig. 4.** Transient thermal response of supply air due to the TE heat pump in cooling mode with electric current 0.5 A at two different mass flowrates, 3 and 5 g/s. The time constants are found to be 1.1 ~ 1.2 seconds for both flowrates.

Temperature distribution inside the system reveals useful information about the heat transfer through the model system. Fig. 5 shows the temperature distribution in the vertical direction in the 1D half system model described in Fig. 3(b) in the cooling mode. The left side of the figure is the top side of the system, i.e., the supply air side, and the right side is the bottom of the system, i.e., the exhaust air side. In the figure, mass flowrate was fixed at 3 g/s and the results for two currents, 1 A and 2 A, are shown. The temperatures measured experimentally are shown as symbols while the theoretical calculations are shown as lines. As seen in the figure, the theory matches the experimental data reasonably well for both currents. As the current increases from 1 A to 2 A, the supply air temperature is reduced from 22.4 to 22.1 °C due to the enhanced cooling. Although a small temperature drop was achieved from 1 A to 2 A, the internal temperatures in the TE module are substantially changed. At 1 A, the temperature difference created between the two ends of TE legs, $(T_{TE,ea} - T_{TE,sa})$ was 12.6 °C (= 29.8 – 17.2 °C). That difference increased substantially to 25.5 °C (= 40.2 – 14.7 °C). Although the Peltier cooling power is doubled ($\propto I$) when current increases from 1 to 2 A, the Joule heating is four times larger as it is proportional to $I^2$, which in turn increases the hot-side temperature, $T_{TE,\text{ea}}$, much more than the reduction in the cold-side temperature, $T_{TE,\text{sa}}$.

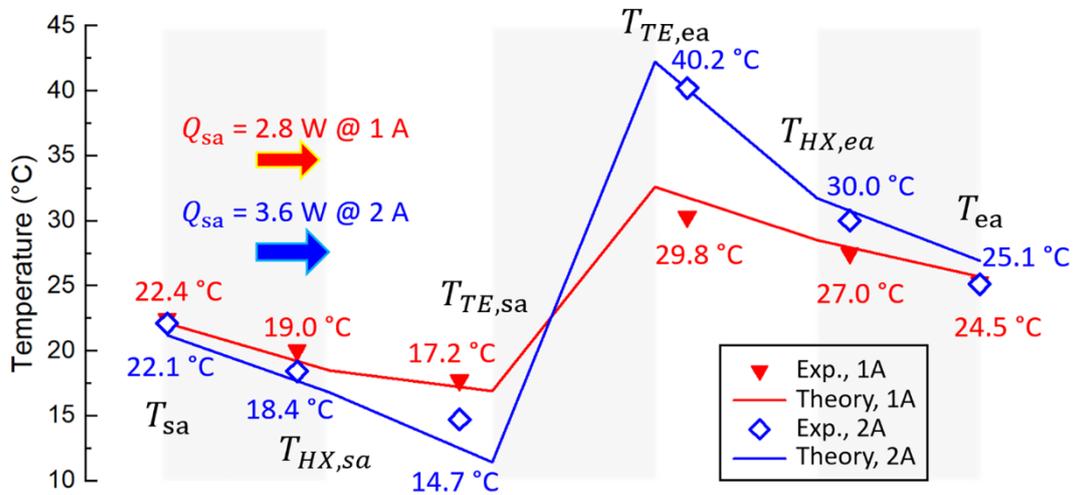

**Fig. 5.** Experimental and theoretical temperature distribution in the vertical direction in the 1D half-system model at currents 1 A and 2 A at fixed mass flowrate 3g/s. Lines are theoretical tempeatures and symbols are experimental temperatures. Results at 1 A are shown in red, and 2 A in blue. Distances along the horizontal axis are not to scale.

At the cold side of TE legs, conduction heat comes in from both directions due to the negative slope of temperature from both directions. The negative slope of temperature from the supply air to the TE module drives the cooling of supply air. The negative slope of temperature from the hot side of TE module reduces the cooing of supply air because it increases the cold-side temperature. The Peltier cooling at that position must be large enough to overcome the reverse conduction heat from the hot side. Due to the increased slope of temperature from supply air to the TE module at increased current, the cooling power, $Q_{\text{sa}}$, increased from 2.8 to 3.6 W as current increased from 1 to 2 A. The cooling power increase is not directly proportional to the current increase, due to the detrimental Joule heating and reverse conduction. Note that $Q_{\text{sa}}$ is only for one TE module, thus it is a half of the total pumped heat by the double-sided cooling scheme.

Fig. 6 presents both the cooling and heating performance of our heat pump system. The temperature change of supply air and the COP have been obtained as a function of electric current input for various mass flowrates. The experimental data were reasonably well fitted by the theory for all data points. In the cooling mode, the temperature drop initially increased with increasing current and then reached the peak values before going down at higher currents. This is due to the increased Joule heating at high currents, following $I^2$, while the Peltier cooling is only linearly increased with current. As the flowrate increases, the temperature drop decreases over the entire current range because their inverse relationship as shown in Eq. (1). Note that $\Delta T_{sa}$ and $\dot{m}_{sa}$ are not exactly inversely proportional to each other at fixed current as shown in Fig. 6 because the total pumped heat also changes due to the change of temperature at various points inside the TE system.

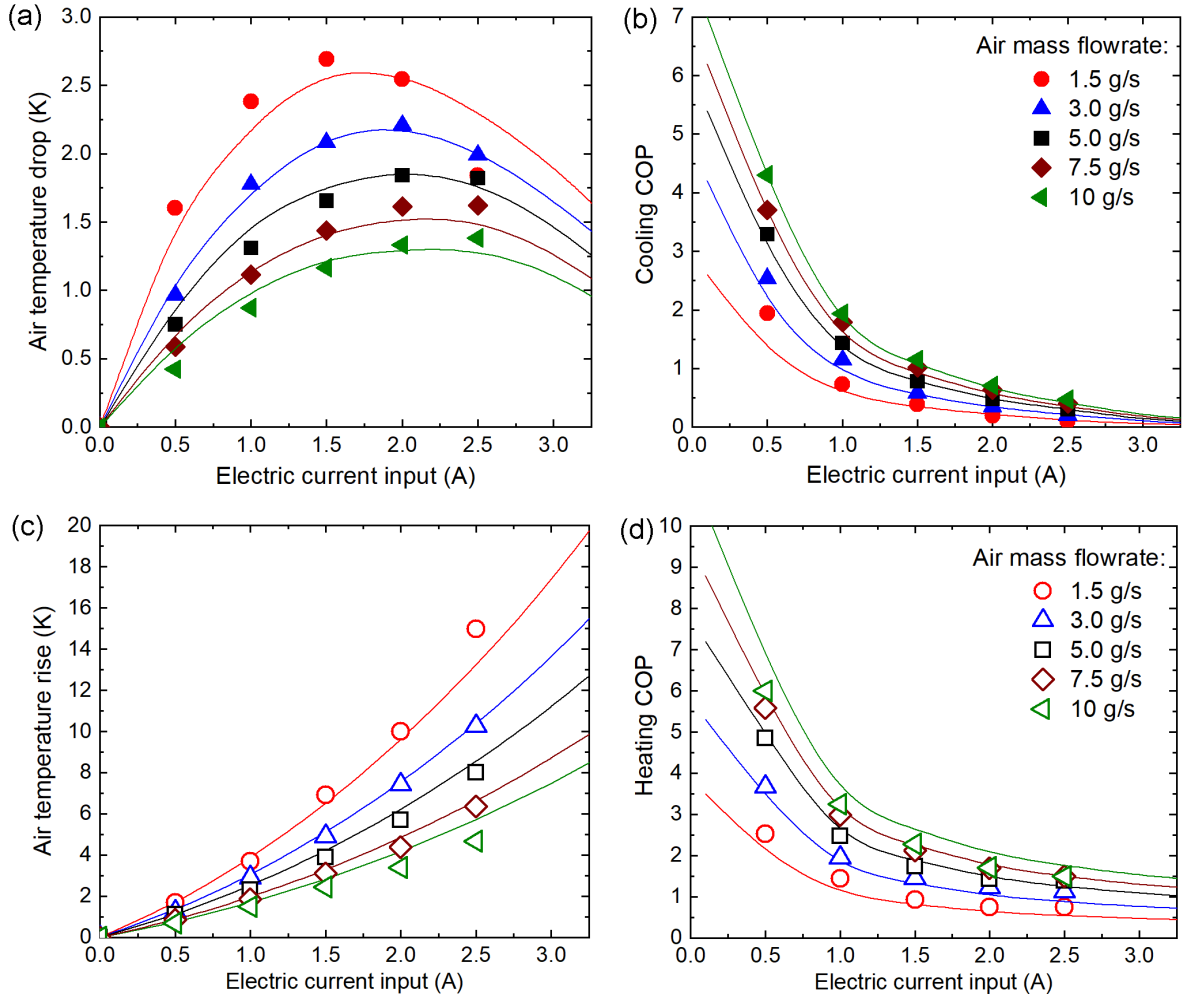

**Fig. 6.** (a)-(b) Cooling performance of the TE heat pump system: (a) Temperature drop of supply air and (b) the cooling COP as a function of electric current for varying mass flowrates. (c)-(d) Heating performance of the TE heat pump system: (c) Temperature rise of supply air and (b) the heating COP as a function of electric current for varying mass flowrates. Symbols are experimental data and curves are theoretical fitting.

The cooling COP gradually increases as the current decreases as previously predicted by our theory [3], proving that the TE modules work more efficiently at lower cooling power. However, if the cooling power is too low, the temperature drop can be too small, potentially falling to the level of parasitic heat losses existing in the system. Too many TE modules would also be needed in series to meet the requirements for temperature drop. Thus, we kept the temperature drop higher than ~ 0.4 K. In this condition, the maximum cooling COP of 4.4 was obtained at 10 g/s flowrate with current at 0.5 A. Note that the COP increased with increasing flowrate at a fixed current. This is due to the increased heat transfer coefficient at higher flowrate, which reduces the overall external thermal resistance in the TE heat pump, and thus increases the cooling power at the same current. Again, higher flowrate would reduce the temperature drop. Mass flowrates ~ 10 g/s or small can achieve such a high COP with reasonably high temperature cooling.

The heating COP shows similar trends with flowrate and current. For $\Delta T_{sa} > \sim 0.7$ K in heating mode, we find that the maximum heating COP of 6.0 is achieved at 0.5 A current and 10 g/s mass flowrate. As current increases or flowrate decreases, the heating COP is reduced. Unlike the cooling mode, the temperature rise of supply air in the heating mode does not show a peak point, but continuously increases with increasing current. This is because the Joule heating is added up to the total heating at the supply air side in heating mode, while Joule heating is subtracted from the total cooling in cooling mode. Temperature increases more rapidly at higher currents because of the rapid increase in Joule heating following $I^2$. Because the Joule heating is added up to the Peltier heating, the maximum COP for heating can be higher than the maximum COP for cooling.

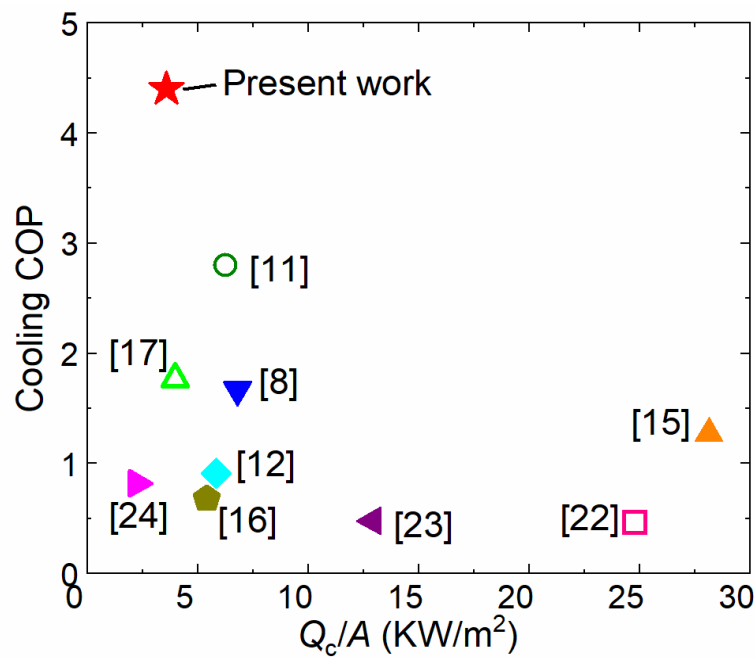

**Fig. 7.** Comparison of cooling COP vs. cooling power density ($Q_c/A$) from literature and this work. Numbers indicate the reference numbers.

Our cooling COP is relatively high compared to previously reported experimental COP values from literature for air-air TE cooling. Fig. 7 summarizes literature COP values vs. their cooling power density ($Q_c/A$). [8, 11, 12, 15, 16, 17, 22, 23, 24] There is a clear trend of a trade-off relationship between COP and cooling power density. Since we aimed to achieve a high COP with optimal TE design in this work, it came with a cost of cooling power density. Targeting at a high COP is beneficial in terms of improving energy efficiency. A low cooling power can be overcome by increasing area or using a greater number of TE modules despite the low power density. This means the increase in the installation cost, but the operation cost can remain low due to the high COP.

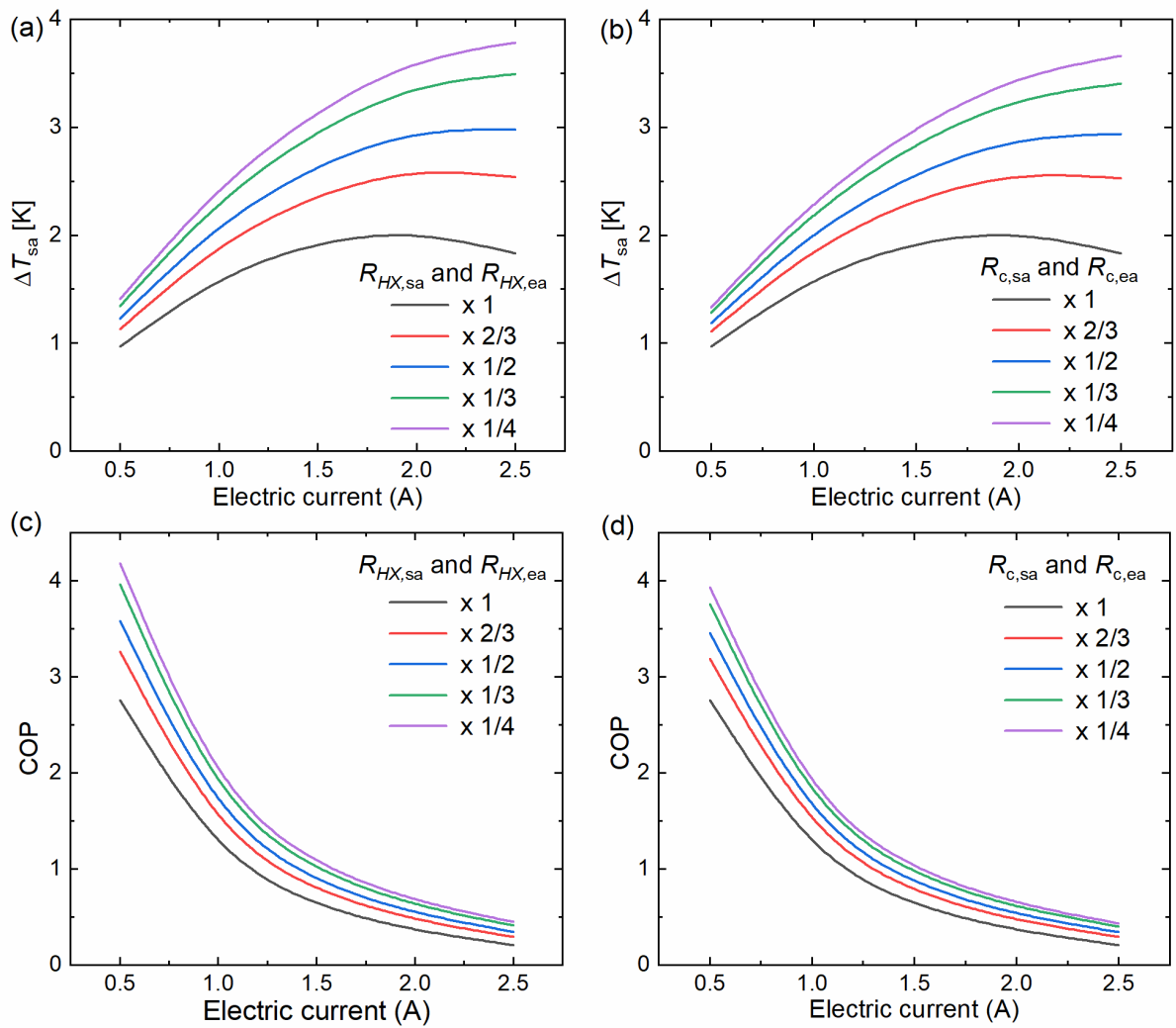

**Fig. 8.** (a)-(b) Calculated air cooling ($\Delta T_{sa}$) and (c)-(d) calculated cooling COP as a function of electric current. (a) and (c): with varying the heat exchanger thermal resistances at both heat exchangers ($R_{HX,sa}$ and $R_{HX,ea}$). (b) and (d): with varying interface resistances at both sides ($R_{c,sa}$ and $R_{c,ea}$). Mass flowrate is fixed at 3 g/s. Varying multiplication factors are given in the legends.

We also theoretically studied the effects of the external thermal resistances in COP and temperature change of supply air in cooling mode. As shown in Fig. 8, the heat exchanger thermal resistances, $R_{HX,sa}$ and $R_{HX,ea}$, as well as the interface conduction thermal resistances, $R_{c,sa}$ and $R_{c,ea}$, were reduced at both sides with several different multiplication factors less than 1 to see the corresponding variation of COP and temperature drop. It is found that both the COP and the temperature drop can be enhanced by more than ~50 % by reducing these external resistances by a factor of ¼. The effects of the heat exchanger resistances and the interface resistances are similar as shown in the figure indicating that their contributions are similar in the final performance.

Reducing the heat exchanger resistance is possible by increasing the surface area of the heat exchanger. In this work, we used simple flat-plate fin heat exchangers, but a much larger surface area, thus a much smaller heat exchanger resistance is possible, for instance, using high-porosity metal foams [25]. Note that metal foams can reduce the heat exchanger resistance substantially only at the expense of increased pressure drop. Increased pressure drop can be translated to an increased pump power to maintain the air flowrate.

Reducing the interface resistance is possible by enhancing the thermal contact or reducing the thicknesses of the ceramic plate of the TE module and the heat exchanger base plate. In the latter, however, the spreading resistance can be increased. [26] Further design optimization is necessary in this regard.

## 5. Conclusions

In short, we experimentally demonstrate high COPs in both cooling and heating modes with optimized TE heat pump system design. We find that such high COPs can be obtained at optimal flowrate and electric current input. A maximum cooling COP of 4.4 and a maximum heating COP of 6.0 were obtained experimentally. Our theoretical analysis indicates that this cooling/heating performance can be further enhanced by reducing the external thermal resistances such as the convection thermal resistance at the heat exchangers and the interface thermal resistances. Our design optimization is based on improving COPs with a relatively low cooling/heating power density. Our strategy is that using a greater number of TE modules in series and in parallel to scale up the cooling power and the flowrate to desired levels while individual TE modules deal with a small cooling power for a high system COP. This combined experimental and theoretical study is expected to be beneficial to push the limits of thermoelectric heat pump systems towards the acceptable and competitive levels at the market in comparison to those of the conventional vapor-compression systems.

## ACKNOWLEDGEMENT

This work was supported by the University of Cincinnati, Office of Research, via the Collaborative Research Advancement Pilot Grants program.